\documentclass{elsart}

\usepackage{amssymb}
\usepackage{graphicx}
\usepackage{hyperref}

\newcommand{\be}{\begin{equation}}
\newcommand{\ee}{\end{equation}}
\newcommand{\bea}{\begin{eqnarray}}
\newcommand{\eea}{\end{eqnarray}}
\newcommand{\eu}{{\rm e}}
\newcommand{\ii}{{\rm i}}
\newcommand{\de}{{\displaystyle\rm\mathstrut d}}
\newcommand{\Pf}{{\rm Pf}}

\begin{document}

\begin{frontmatter}



\title{Emptiness Formation Probability for the Anisotropic XY Spin
Chain in a Magnetic Field}


\author {Alexander G. Abanov, Fabio Franchini}
\address{Physics \& Astronomy Department,}
\address{State University of New York,}
\address{Stony Brook, New York 11794-3800}

\begin{abstract}
We study an asymptotic behavior of the probability of formation
of a ferromagnetic string (referred to as EFP) of length $n$ in
a ground state of the one-dimensional
anisotropic XY model in a transversal magnetic field as
$n\to \infty$.
We find that it is exponential everywhere in the phase diagram 
of the XY model except at the critical lines where the spectrum
is gapless.
One of those lines corresponds to the isotropic XY model where
EFP decays in a Gaussian way, as was shown in
Ref.\cite{shiroishi}.
The other lines are at the critical value of the magnetic field.
There, we show that EFP is still exponential but acquires a
non-trivial power-law prefactor with a universal exponent.
\end{abstract}

\end{frontmatter}

\section{Introduction}

Emptiness Formation Probability (EFP) is a special correlation function
which plays an important role in the theory of integrable models
\cite{korepin93}.
In the case of spin-$\frac{1}{2}$ chains at zero temperature it is
defined as 
\be
   P(n) \equiv
   \left\langle  \prod_{j=1}^n {1 - \sigma_j^z \over 2}  \right\rangle
   \label{EFPDef}
\ee
where $\sigma_i^{x,y,z}$ are Pauli matrices, the integer $j$ labels
the sites of the 1d lattice, and the average in (\ref{EFPDef}) is
performed over the ground state of the Hamiltonian of the spin chain.
EFP measures the probability that $n$ consecutive spin sites are
all found aligned downward in the ground state.
The asymptotic behavior of $P(n)$ at large $n$ is, probably, the most
interesting one from the physical point of view, as it is believed that
it measures the tendency of the system to a phase separation.

There has been considerable progress in finding this asymptotic behavior
in the case of the integrable XXZ spin chain in critical regime
\cite{korepin93,korepin94,shiroishi,stroganov1,KMST_0-2002,KMST_gen-2002,KLNS-2002}.
At finite temperature the asymptotic behavior of EFP is exponential
\cite{korepin93} $P(n)\sim e^{-\gamma n}$, while at zero temperature in
critical regime it is Gaussian $P(n)\sim e^{-\alpha n^{2}}$. 
The qualitative argument in favor of a Gaussian decay of EFP at zero
temperature was given in Ref. \cite{abanovkor} in the framework of a
field theoretical approach.
There, it was argued that an optimal fluctuation of the field
corresponding to EFP in the critical model will have (very roughly)
the form of a round droplet in space-time with area $\sim n^{2}$ and
will have an action proportional to $n^{2}$.
The asymptotic form of $P(n)$  is defined by the action of this optimal
configuration and is proportional to $P(n)\sim e^{-\alpha n^{2}}$,
which explains the Gaussian decay of EFP at zero temperature.
Similarly, at finite temperature $T$ in the limit $n\gg 1/T$ the droplet
becomes rectangular (one dimension $n$ is replaced by the inverse
temperature $1/T$) and the action cost is proportional to $n$,
giving $P(n)\sim e^{-\gamma n}$.
This argument is tied to the criticality of the theory.
Indeed, for noncritical theory one should not expect the same scaling of
time and space dimensions of an optimal configuration.
It is, therefore, very interesting to see how the asymptotic behavior
of EFP is related to the criticality of a theory. 

\begin{figure}
  \includegraphics[width=\columnwidth]{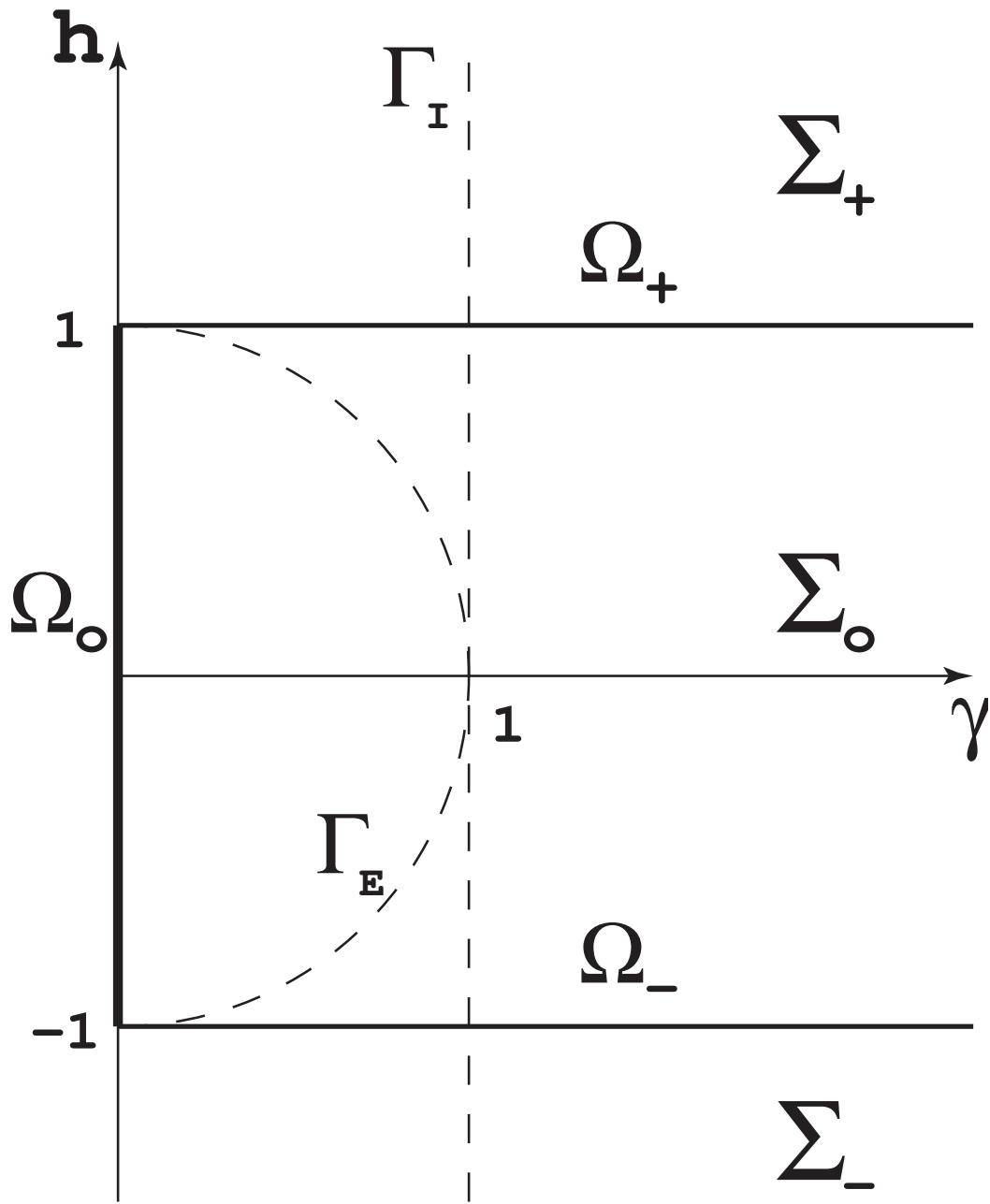}
\caption{Phase diagram of the XY Model. }
   \label{phasespace}
\end{figure}

In this paper we examine the relation between an asymptotic behavior
of EFP and criticality using the example of the Anisotropic XY spin-1/2
chain in a transverse magnetic field $h$
\be
   H = \sum_{j} \left[
   \left( {1 + \gamma \over 2} \right) \sigma_j^x \sigma_{j+1}^x +
   \left( {1 - \gamma \over 2} \right) \sigma_j^y \sigma_{j+1}^y \right]
   - h \sum_{j} \sigma_j^z
   \label{spinham}
\ee
This model has been solved in \cite{LSM-1961} in the case of zero
magnetic field and in \cite{mccoy} in the presence of a magnetic field.
The phase diagram is depicted in Fig. \ref{phasespace}.
The phase diagram of the XY model has obvious symmetries
$\gamma\to -\gamma$ and $h\to -h$.
However, the latter one is broken by EFP itself.
Therefore, we show only the part of the diagram corresponding to
$\gamma \ge 0$.
The phase diagram has both critical and non-critical regimes.
The first critical regime -- the Isotropic ($\gamma=0$)  XY model --
is labeled $\Omega_0$.
There is another critical line $h=1$ which is labeled $\Omega_+$ where
excitations are gapless (and similarly line $\Omega_-$ at $h=-1$).
The rest of the phase diagram is non-critical with finite gap for
excitations and is divided into domains $\Sigma_+$, $\Sigma_0$, and
$\Sigma_-$ for $h>1$, $-1<h<1$, and $h<-1$ respectively.
Fig. \ref{phasespace} includes the line $\gamma=1$ ($\Gamma_I$)
corresponding to an Ising model in transverse magnetic field and the
line $\gamma^2+h^2=1$ ($\Gamma_E$) on which the wave function of the
ground state is factorized into a product of single spin states
\cite{shrock}. 

\section{The Emptiness Formation Probability in the XY Model}
\label{XYModel}

It was shown in Refs. \cite{LSM-1961,mccoy} that spin-spin correlation
functions in the XY model can be represented as determinants of some
Toeplitz matrices so that one can find their asymptotic behavior using
known theorems from the theory of Toeplitz matrices.
This approach has been applied to EFP for the isotropic ($\gamma=0$) XY
model in magnetic field in Ref. \cite{shiroishi}.
In this section we (slightly) generalize the derivation of Ref.
\cite{shiroishi} to the case of $\gamma\neq 0$ and represent EFP as a
determinant of a Toeplitz matrix.

We follow the standard prescription \cite{LSM-1961} and reformulate the
Hamiltonian (\ref{spinham}) in terms of spinless fermions $\psi_i$ by
means of a Jordan-Wigner transformation:
\bea
   \sigma_j^+ & = &
   \psi_j^\dagger \eu^{\ii \pi \sum_{k<i} \psi_k^\dagger \psi_k} =
   \psi_j^\dagger \prod_{k<j}
   \left( 2 \psi_k^\dagger \psi_k - 1 \right), \\
   \sigma_j^z & = & 2 \psi_j^\dagger \psi_j - 1,
\eea
where, as usual, $\sigma^\pm = (\sigma^x \pm i \sigma^y)/2$.
In Fourier components $\psi_j = \sum_q \psi_q \eu^{\ii q j}$ we have:
\be
   H = \sum_q \left[
   2 \left( \cos \! q - h \right) \psi_q^\dagger \psi_q
   + \ii \gamma \sin \! q \: \psi_q^\dagger \psi_{-q}^\dagger
   - \ii \gamma \sin \! q \: \psi_{-q} \psi_q \right].
   \label{spinlessham}
\ee
The Bogoliubov transformation
\be
   \chi_{q} = \cos \! {\vartheta_q \over 2} \: \psi_{q} 
   + \ii \sin \! {\vartheta_q \over 2} \: \psi_{-q}^\dagger
   \label{bogtrans}
\ee
with ``rotation angle'' $\vartheta_q$ 
\be
   e^{i\vartheta_q}  = \frac{1}{\varepsilon_q}(\cos q-h+i\gamma\sin q),
\ee
brings the Hamiltonian (\ref{spinlessham}) to a diagonal form
$\sum_q \varepsilon_q \chi_q^\dagger\chi_q$ with the quasiparticle spectrum
\be
   \varepsilon_q = \sqrt{ \left( \cos q - h \right)^2
   + \gamma^2  \sin^2 q}.
   \label{spectrum}
\ee
The fermionic correlation functions are also easy to obtain from
(\ref{spinlessham}).
In the thermodynamic limit they read \cite{LSM-1961,mccoy} 
\bea
   F_{jk} &\equiv& \ii\langle \psi_j \psi_k \rangle
   = - \ii\langle \psi_j^\dagger \psi_k^\dagger \rangle
   = \int_{-\pi}^\pi {\de q \over 2\pi}\; \frac{1}{2}\sin\vartheta_q
   \eu^{\ii q (j-k)} 
  \label{F} 
\\
   G_{jk} &\equiv& \langle \psi_j \psi_k^\dagger \rangle
   =   \int_{-\pi}^\pi {\de q \over 2 \pi}\; \frac{1+\cos\vartheta_q}{2}
   \eu^{\ii q (j-k)} .
 \label{G}
\eea

We express EFP as the fermionic correlator \cite{shiroishi}
\be
   P(n) = \langle \prod_{j=1}^n \psi_j \psi_j^\dagger \rangle.
   \label{expect}
\ee
and using Wick's theorem on the r.h.s of (\ref{expect}), we obtain
\be
    P(n) =  \Pf( {\bf M} ) \equiv \sqrt{\det( {\bf M} )},
\ee
where $\Pf( {\bf M} )$ is the {\it Pfaffian} of the $2n \times 2n$
skew-symmetric matrix ${\bf M}$ of correlation functions
\be
   {\bf M} = \pmatrix{  {-\ii\bf F} &
                        {\bf G}  \cr
                        - {\bf G}  &
                        \ii{\bf F} \cr
                     }.
\ee
Here ${\bf F}$ and ${\bf G}$ are $n\times n$ matrices with matrix
elements given by $F_{jk}$ and $G_{jk}$ from (\ref{F},\ref{G})
respectively.

We perform a unitary transformation
\be
   {\bf M'} = {\bf U M U}^{\dagger}
   = \pmatrix { 0 & {\bf S_n} \cr -{\bf S_n}^\dagger & 0 \cr}, \qquad
   {\bf U} = {1 \over \sqrt{2}} \pmatrix { {\bf I} & {\bf -I} \cr
                                           {\bf I} & {\bf I} \cr},
\ee
where ${\bf I}$ is the $n\times n$ unit matrix and ${\bf S_n}={\bf 
G}+\ii{\bf F}$ and ${\bf S_n}^\dagger={\bf G}-\ii{\bf F}$. This allows 
us to calculate the determinant of ${\bf M}$ as
\be
   \det ( {\bf M} ) = \det ( {\bf M'} ) =
   \det ( {\bf S_n} ) \cdot \det ( {\bf S_n}^\dagger )
   = \left|\det ( {\bf S_n} ) \right|^2.
   \label{detM}
\ee
The matrix ${\bf S_n}$ is a $n\times n$ Toeplitz matrix (its matrix 
elements depend only on the difference of raw and column 
indices \cite{basor}).
The generating function $\sigma(q)$ of a Toeplitz matrix is defined by
\be
    ({\bf S_n})_{jk}=\int_{-\pi}^\pi {d q \over 2 \pi}\; \sigma (q)
    \eu^{\ii q(j-k)}
 \label{Tg}
\ee
and can be found from (\ref{F},\ref{G}) as
\be
   \sigma(q) = {1 \over 2}\left( 1 + \eu^{\ii\vartheta_q} \right)
   = \frac{1}{2} 
    +\frac{\cos q -h +i\gamma\sin q}{2\sqrt{(\cos q-h)^2+\gamma^2\sin^2q}}.
   \label{genfunc}
\ee

Thus, the problem of calculation of EFP  
\be
   P(n) =  \left|\det ( {\bf S_n} ) \right|,
  \label{PnX}
\ee
is reduced (exactly) to the calculation of the determinant of the
$n\times n$ Toeplitz matrix ${\bf S_n}$ defined by the generating
function (\ref{Tg},\ref{genfunc}).
The representation (\ref{PnX}) is exact and valid for any $n$.
In our study we are interested in finding an asymptotic behavior at
large $n \to \infty$.

In the following section we apply known theorems on the asymptotic
behavior of Toeplitz determinants to extract the corresponding
asymptotes of $P(n)$ at $n\to \infty$.
The reader who is interested in results can skip the following
section and go directly to Sec.\ref{Conclusions} where we present
and discuss our results.

\section{The results for the asymptotic behavior of EFP}
\label{results}

The asymptotic behavior of EFP at $n\to\infty$ is exactly related to the
asymptotic behavior of the determinant of the corresponding Toeplitz
matrix (\ref{Tg},\ref{genfunc},\ref{PnX}) and can be extracted from known
theorems and conjectures in the theory of Toeplitz matrices.
These types of calculations have been done first in
\cite{LSM-1961,mccoy} for spin-spin correlation functions.
It is well known that the asymptotic behavior of the determinant of a
Toeplitz matrix as the size of the matrix goes to infinity strongly
depends upon the zeros of the generating function of the matrix and
upon its phase.

In the case when the generating function $\sigma(q)$ has only isolated
singularities (zeros and phase discontinuities) the most general 
known result for the asymptotic behavior of a
Toeplitz determinant is called {\it generalized Fisher-Hartwig conjecture}
(gFH) and is due to Basor and Tracy \cite{basor}.
The result can be written
as 
\be
    P(n) \sim \sum_{i} E_{i} n^{-\lambda_{i}} e^{-\beta_{i} n},
 \label{FHsymb}
\ee
where the decay rates $\beta_{i}$, the exponents $\lambda_{i}$, and the
coefficients $E_{i}$ are known functionals of $\sigma(q)$ \cite{basor}.
The original {\it  Fisher-Hartwig conjecture} (FH)
\cite{FisherHartwig-1968} consists of just one term of the sum in 
(\ref{FHsymb}). We apply gFH to the generating function
(\ref{genfunc}) and extract the asymptotic behavior of $P(n)$.
All necessary theorems on the asymptotic behavior of Toeplitz 
determinants can be found in
Ref.\cite{wu,FisherHartwig-1968,widomsing,basor,Ehrhardt-1997,widomsupp}.
For the most recent review of gFH see also Ref.\cite{Ehrhardt-2001}.

In some parts of the phase diagram we have to rely on ``conjectures'',
i.e., results which are not proven yet.
Therefore, it is prudent to look for some independent ways of checking
those results.
We did some numeric calculations on these Toeplitz determinants and the
results of these calculations support the analytic results presented
below \cite{AF}.
Also, we used some limiting cases where EFP or an asymptotic behavior
of Toeplitz determinant can be obtained in an independent way to
confirm our conclusions.
E.g., on the line $\Gamma_E$ EFP can be calculated exactly and the
asymptotic behavior of the result is the same as the one obtained
using FH for the whole region $\Sigma_0$ of the phase diagram.

In the rest of this section we present the results for the asymptotic
behavior of $P(n)$ in different regions of the phase diagram of the XY
model (Fig. \ref{phasespace}).
In this letter we concentrate on the calculation of $\lambda$ and
$\beta(h,\gamma)$ leaving our findings for the coefficient
$E(h,\gamma)$ as well as the results of our numeric calculations for a
future, more extended publication \cite{AF}.

\subsection{The non-critical theory ($\Sigma_\pm$ and $\Sigma_0$)}

\subsubsection{The region $\Sigma_-$}

In this region ($\gamma\neq 0$,$h<-1$) the generating function
(\ref{genfunc}) is nonzero for all $q$ and the FH conjecture is
reduced to a (rigorously proven) {\it Strong Szeg\"o Limit Theorem}.
It gives
\be
    P(n)\sim E(h,\gamma) e^{-\beta(h,\gamma) n}
 \label{pnc}
\ee
with
\bea
   \beta(h, \gamma) & = &
   - \int_{-\pi}^\pi {\de q \over 2\pi}\; \log  \left| \sigma(q) \right|
   \nonumber \\ & = &
   - \int_0^\pi {\de q \over 2\pi}\;
   \log \left[ {1 \over 2} \left( 1 + {\cos q -h \over \sqrt{
   \left( \cos q - h \right)^2 + \gamma^2  \sin^2 q} } \right) \right].
   \label{betagh} 
\eea
The integral in (\ref{betagh}) is convergent for all $h$ and all
$\gamma\neq 0$ and $\beta(h,\gamma)$ is a continuous function of its
parameters.
In fact, it turns out (see below) that this expression for
$\beta(h,\gamma)$ describes correctly the decay rate of EFP everywhere
in the phase diagram except for $\gamma=0$.
However, both pre-exponential factors and the mathematical status of
the results are quite different in different parts of the phase diagram
(see below). 

In Figure \ref{betagraph}, $\beta(h,\gamma)$ is plotted as a function
of $h$ at several values of $\gamma$.
One can see that $\beta(h,\gamma)$ is continuous but has a weak
(logarithmic) singularity at $h=\pm 1$.
This is obviously one of the effects of the criticality of the model on
the asymptotic behavior of EFP.

\begin{figure}
  \includegraphics[width=\columnwidth]{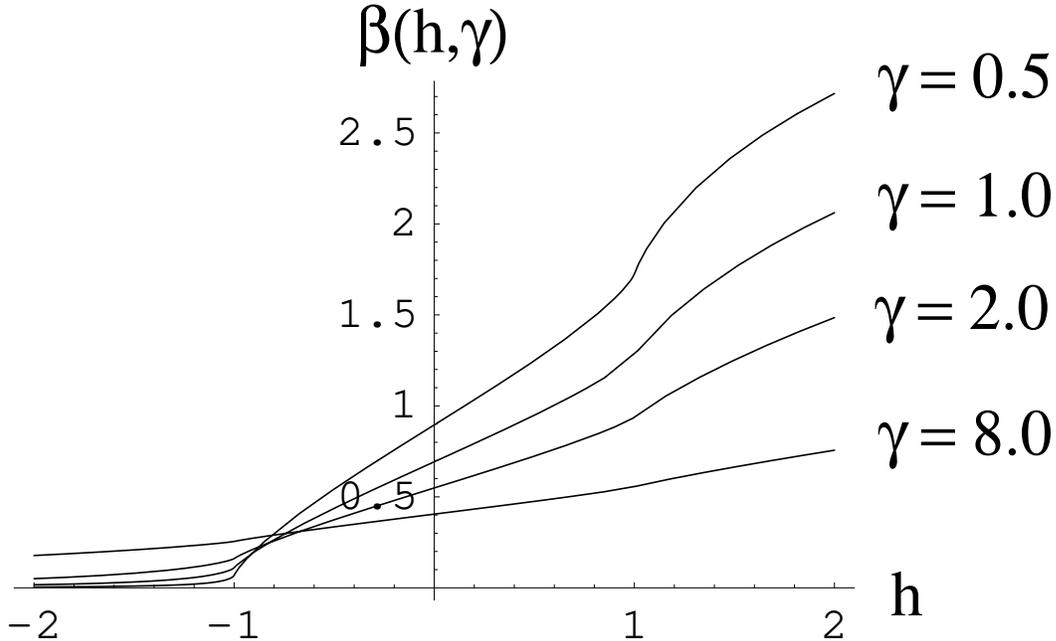}

\caption{Plot of the decay factor $\beta$ as a function of the
parameters $\gamma$ and $h$. The function diverges for $\gamma = 0$
and is continuous for $h = \pm 1$ (although has a weak singularity at $h=\pm 1$).}

   \label{betagraph}
\end{figure}

\subsubsection{The region $\Sigma_0$}

In $\Sigma_0$ ($\gamma\neq 0$,$-1<h<1$) the generating function
$\sigma(q)$ vanishes at $q = \pm \pi$ and acquires an overall phase of
$\pi$ as $q$ goes from $-\pi$ to $\pi$.
The Szeg\"o theorem is not applicable anymore.
One has to rely on the FH conjecture (which, in fact is proven for this
case -- the case of one singularity \cite{Ehrhardt-1997}) to derive the
result, which is identical to (\ref{pnc},\ref{betagh}).
The expression for the coefficient $E(h,\gamma)$ is different from the
one in $\Sigma_-$ but its form is beyond the scope of this paper.

\subsubsection{The region $\Sigma_+$}

In $\Sigma_+$ ($\gamma\neq 0$, $h>1$) the generating function $\sigma(q)$  
vanishes at the points $q = 0$ and
$q = \pm \pi$, and its phase has jumps at these points.
In this case an application of FH leads to some ambiguity, e.g., one
obtains two values for $\beta(h,\gamma)$: $\beta_1=\beta$ and 
$\beta_2=\beta+\pi i$ with $\beta$ from (\ref{betagh}). 
This ambiguity is resolved in the use of the gFH which gives EFP as a
sum of two terms so that both values of $\beta$ are used.

We obtain
\be
   P(n) \sim {\it E}(h,\gamma)
   \left[ 1 + (-1)^n A(h,\gamma) \right] \: \eu^{- n \beta(h, \gamma)}.
   \label{oscbehavior}
\ee
Thus, in this region we again have an exponential asymptotic decay of 
EFP but, the constant prefactor is different for even and odd $n$.

\subsubsection{The line $\Gamma_2$: an exact calculation}

Before proceeding to the analysis of EFP on the critical lines let us 
check our results (\ref{pnc},\ref{betagh}) on the special line\footnote{We 
are grateful to Fabian Essler who suggested us to check our results 
on this special line and pointed out the reference \cite{shrock} to 
us.} in the phase diagram defined by
\be
   h^2 + \gamma^2 = 1.
   \label{trajectory}
\ee
It was shown in Ref.\cite{shrock} that on this line the ground state 
is a product of single spin states and is given by
\be
   \vert G \rangle = \prod_j \vert \theta, j \rangle = \prod_j
   \left[ \cos \left( {\theta \over 2} \right) | \uparrow, j \rangle +
   (-1)^j \sin \left( {\theta \over 2} \right) | \downarrow, j \rangle
   \right],
   \label{groundstate}
\ee

where $| \uparrow, j \rangle$ is an up-spin state at the lattice 
site $j$ etc.
One can directly check that the state 
(\ref{groundstate}) is an eigenstate of (\ref{spinham}) if the value 
of the parameter $\theta$ is
\be
    \cos^2 \theta  =  \frac{1-\gamma}{1 + \gamma}
\ee
and (\ref{trajectory}) is satisfied. It is also easy to 
show \cite{shrock} that this state is, in fact, the ground state of 
(\ref{spinham}).

The probability of formation of a ferromagnetic string in the state 
(\ref{groundstate}) is obviously
\be
   P(n) = \sin^{2n} \left( {\theta \over 2} \right) 
    = \left( {1 \over 2} - {1 \over 2} 
    \sqrt{ 1 - \gamma \over {1 +\gamma} } 
\right)^n
 \label{pnexact}
\ee
which is an exact result on the line (\ref{trajectory}).
The value of  $\beta(h,\gamma)$ which immediately follows from this
exact result is
\be
   \beta(h=\sqrt{1-\gamma^2},\gamma) =
   - \log\left( {1 \over 2} - {1 \over 2} 
   \sqrt{ 1 - \gamma \over {1 +\gamma} } \right) 
 \label{betaexact}
\ee
Indeed, one can easily check that the integral in the general formula
for the decay rate (\ref{betagh}) under the condition
(\ref{trajectory}) gives precisely (\ref{betaexact}).
In fact, the Toeplitz matrix generated by (\ref{genfunc}) becomes
triangular on the line (\ref{trajectory}) with diagonal matrix
element $(S_n)_{jj} = \sin^2(\theta/2)$ and the determinant of
${\bf S_n}$ is exactly (\ref{pnexact}).

\subsection{The critical line $\Omega_0$ ($\gamma = 0$)}
\label{gammazero}

The case $\gamma = 0$, corresponding to the Isotropic XY Model, has
been extensively studied in Ref.\cite{shiroishi}.
For $\gamma=0$ the generating functions (\ref{genfunc}) is reduced
to the one found in \cite{shiroishi}.

For $h < 1$, the generating function $\sigma(q)$ has a limited
support: to find the asymptotic behavior of the determinant of the
Toeplitz matrix one can apply Widom's Theorem \cite{widomsupp} and
obtain \cite{shiroishi}:
\be
   P(n) \sim
   2^{5 \over 24} \eu^{3 \zeta'(-1)} (1-h)^{- {1 \over 8}}
   n^{- {1 \over 4}} \left( {1 + h \over 2} \right)^{n^2 \over 2}
 \label{pnsh}
\ee 
and we see that in this case, EFP decay as a Gaussian with an
additional power-law pre-factor.\footnote{Notice, that the formula
(\ref{pnsh}) was in fact obtained in \cite{dysmehta} in the context
of unitary random matrices (which are known to be equivalent to free
fermions).}

For $|h| > 1$, the theory in not critical anymore and the ground state
is completely polarized in $z$ direction, giving a trivial EFP $P(n)=0$
for $h>1$ and $P(n)=1$ for $h<-1$.

\subsection{The critical lines $\Omega_\pm$($h = \pm 1$)}

\subsubsection{The critical line $\Omega_+$($h = 1$)}
\label{ssOmegap}

\begin{figure}
  \includegraphics[width=\columnwidth]{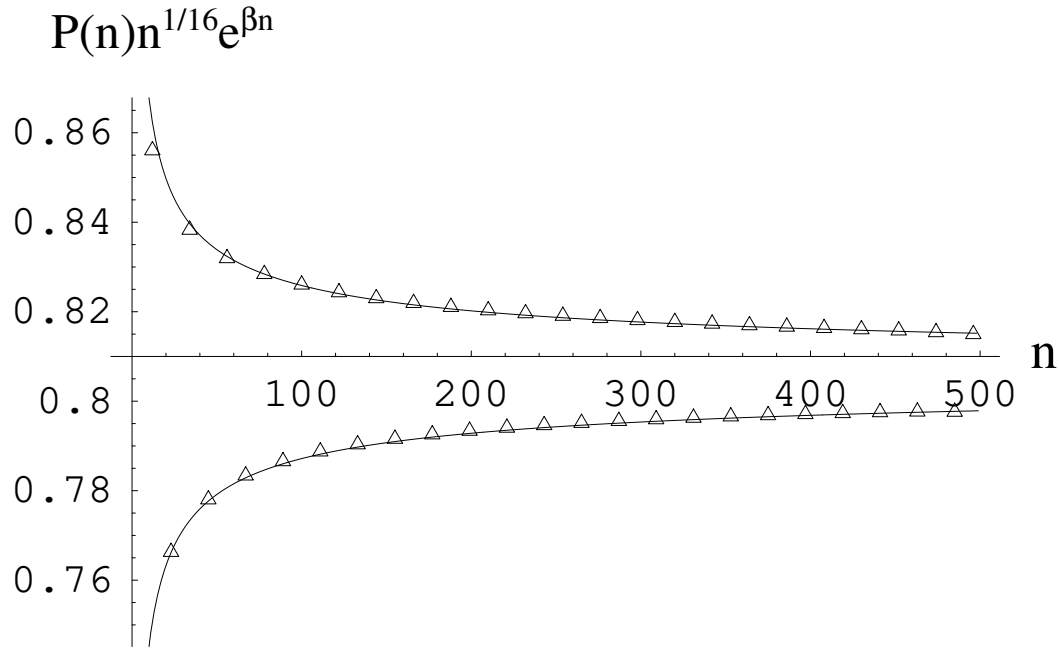}

\caption{Results of the numeric calculation of the Toeplitz determinant
is shown as $P(n) e^{\beta n} n^{1/16}$ vs $n$ at $\gamma =1$, $h=1$.
The value of $\beta=\log 2 +2G/\pi$ with Catalan's constant $G$ is obtained
from (\ref{betagh}).
The solid line is the analytic result $E(1+(-1)^n A/n^{1 \over 2})$ with
$E=0.8065...$, $A=0.2399...$ obtained by fitting at large $n$. To make the
plot more readable we show only every 11th point (for $n=1,12,23,\ldots$).}

   \label{DetPlot}
\end{figure}

For $h=1$ the generating function $\sigma(q)$ vanishes at $q = \pm \pi$
and its phase has jumps at $q = 0,\pi$.
The gFH in this case produces many terms of the form (\ref{FHsymb}).
However, in contrast to the $\Sigma_+$ region, all terms are suppressed
by power law factors of $n$ with respect to the leading one.
The leading term has $\beta$ from (\ref{betagh}) and critical exponent
$\lambda = {1\over 16}$:
\be
   P(n) \sim E(\gamma) n^{- {1 \over 16} } \: \eu^{-n \beta(1, \gamma)}.
 \nonumber
\ee
The next subleading term (faster decay at $n\to \infty$) is given by
\be
   E'(\gamma) (-1)^n n^{- {9 \over 16} } \: \eu^{-n \beta(1, \gamma)}.
 \nonumber
\ee
Although inclusion of the latter (subleading) term is somewhat beyond
even the generalized FH, we write the sum of these two terms as a
conjecture for EFP at $h=1$ and obtain
\be
   P(n) \sim {\it E}(\gamma) \:
   n^{- {1 \over 16} } \left[ 1 + (-1)^n A(\gamma)/n^{1 \over 2} 
   \right] \: \eu^{- n \beta(1, \gamma)}.
 \label{pnomp}
\ee
As this result relies on this unproven conjecture, we present our numeric
results for this case in Figure \ref{DetPlot}.
Indeed, we see that the form (\ref{pnomp}) is in good agreement with numerics.
We conclude that at $h=1$ EFP decays exponentially at $n\to\infty$ but
with an additional power law pre-factor and a damped oscillatory component.

{\it Remark.} It is curious to notice that $1/16$ and $9/16$ remind of the scaling
dimensions of spins $\sigma^x$ and $\sigma^y$.\footnote{See Ref.\cite{mccoy}
where it was shown that power laws are $1/4$ and $9/4$ for $\sigma^x$ and
$\sigma^y$ correlators respectively.}
It looks as if EFP operator (\ref{EFPDef}) among other things has inserted
square roots of spins at the ends of the string.

\subsubsection{The critical line $\Omega_-$($h = -1$)}

For $h=-1$ the generating function $\sigma(q)$ does not vanish but has a
phase discontinuity at $q=\pm \pi$.
We apply FH to obtain
\be
   P(n) \sim {\it E}(\gamma) \:
   n^{- {1 \over 16} }  \: \eu^{- n \beta(-1, \gamma)}.
\ee
and stretching gFH the same way as we did for $h=+1$ we include a subleading term
\be
   P(n) \sim {\it E}(\gamma) \:
   n^{- {1 \over 16} } \left[ 1 +  A(\gamma)/n^{1 \over 2} 
   \right] \: \eu^{- n \beta(-1, \gamma)}.
 \label{pnomm}
\ee
This form is also in good agreement with our numeric results.

\section{Discussion and Conclusions}
\label{Conclusions}

An asymptotic behavior of Emptiness Formation Probability $P(n)$ as 
$n\to\infty$ for the Anisotropic XY model in a transverse magnetic field
as a function of the anisotropy $\gamma$ and the magnetic field $h$
has been studied.
We summarize our results in Table \ref{table1}.
All these results are new except for the isotropic case $\Omega_0$
($\gamma=0$) which was studied in Ref.\cite{shiroishi}.

\begin{table}
\caption{Asymptotic behavior of EFP in the different regimes.
 The exponential decay rate $\beta$ is given by Eq. (\ref{betagh})
 for all regimes.
 The coefficients $E,A$ are functions of $h$ and $\gamma$ as appropriate.}
 \centering
 \label{table1}
   \begin{tabular}{|c|c|c|l|l|l|}
     \hline
     \multicolumn{6}{|c|}{\bfseries EFP for an Anisotropic XY model} \\
     \hline
      Region  &  $\gamma$, $h$  & criticality &  $P(n)$  &  Eq.\#  &  Theorem \\
     \hline
     \hline
      $\Sigma_-$  &  $h<-1$  & No & $ E \eu^{-n \beta}$  &  \ref{pnc}  &  Szeg\"o \\
     \hline
      $\Sigma_0$  &  $-1<h<1$  &  No & $ E \eu^{-n \beta }$  &  \ref{pnc}  &  FH  \\
     \hline
      $\Sigma_+$  &  $h>1$  & No &
       $ {\it E} \left[ 1 + (-1)^n A \right] \: \eu^{-n \beta}$  &  \ref{oscbehavior}  &  gFH \\ 
     \hline
      $\Gamma_2$  &  $\gamma^2+h^2=1$  &  No & $ E \eu^{-n \beta}$  &  \ref{pnexact}  &  Exact \\ 
     \hline
 \hline
      $\Omega_+$  &  $h=1$  &   Yes &
       $ {\it E} \: n^{- {1 / 16} } \left[ 1 + (-1)^n A/\sqrt{n} \: \right] \: \eu^{- n \beta}$  &  \ref{pnomp}  &  gFH  \\ 
     \hline
      $\Omega_-$  &  $h=-1$  & Yes & $ {\it E} \: n^{- {1 / 16} } \left[ 1 +  A/\sqrt{n} \: \right]  \: \eu^{- n \beta}$  &  \ref{pnomm}  &  gFH \\
     \hline
      $\Omega_0$  &  $\gamma=0$,\, $-1<h<1$  & Yes & $ E n^{-1/4}e^{-n^2 \alpha}$  &  \ref{pnsh}  &  Widom  \\
     \hline 
   \end{tabular}
\end{table}

Our main motivation was to study the relation between the criticality of
the theory and the asymptotics of EFP.
Let us look at the results on the critical lines $\Omega_0$ and $\Omega_\pm$.
The Gaussian behavior on $\Omega_0$ is in accord with the qualitative argument
of Ref.\cite{abanovkor}.
However, on $\Omega_\pm$ the decay of EFP is exponential instead of Gaussian,
and apparently contradicts the qualitative picture of Ref.\cite{abanovkor}
(which is reproduced in the introduction to this paper for the sake of completeness).
The reason for this disagreement is that at $h=\pm 1$ the model is critical for the
{\it quasiparticles} $\chi$ defined in (\ref{bogtrans}).
However, we are interested in EFP for the ``original'' Jordan-Wigner fermions $\psi$,
and this correlator has clearly a different expression in terms of $\chi$ than the
simple one (\ref{expect}) which holds at $\gamma =0$.
From the technical point of view, the difference is that in the qualitative argument
in favor of a Gaussian decay of EFP for critical systems there is an implicit
assumption that the density of fermions (or magnetization) is related in a local way
to the field responsible for critical degrees of freedom (free boson field $\phi$).
This assumption is not valid on the lines $h=\pm 1$.
Although the theory is critical on those lines and can be described by some
free field $\phi$, the relation between the magnetization and this field is
highly nonlocal and one can not apply the simple argument of \cite{abanovkor}
to the XY model at $h=\pm 1$. 

Although EFP at the critical magnetic field does not show a Gaussian behavior,
there is an important difference between the asymptotic behavior
of EFP on and off the critical lines.
Namely, a power-law pre-factor appears on all critical lines.
For the XY model it is universal (constant on a given critical line) and takes
values $\lambda=1/4$ for $\gamma=0$ \cite{shiroishi} and $\lambda =1/16$ on the
lines $h=\pm 1$. It would be interesting to understand which operators determine
these particular values of ``scaling dimensions'' of EFP
(see the remark at the end of section \ref{ssOmegap}).

In some parts of the phase diagram we used the so-called {\it generalized
Fisher-Hartwig conjecture} \cite{basor} which is not proven yet.
However, our numeric calculations support our analytical results, see, e.g.,
Figure \ref{DetPlot}.
We note that to the best of our knowledge this is the first physically
motivated example where the original Fisher-Hartwig conjecture fails and its
extended version is necessary.
We also suggested that the gFH could be used to find the subleading corrections
to the asymptotic behavior, as we did for $h = \pm 1$ in (\ref{pnomp},\ref{pnomm}).
From the physical point of view, the oscillations of EFP that we obtain for
$h \ge 1$ from the gFH are related to a spontaneous $Z_2$ symmetry breaking which
occurs at $h=1$.

In conclusion, we notice that it is straightforward to generalize our results for
nonzero temperature.
The only modification is that at $T\neq 0$ the thermal correlation functions must
be used instead of (\ref{F},\ref{G}).
Then, the generating function (\ref{genfunc}) is non-singular everywhere and we
have an exponential decay of $P(n)$ in the whole phase diagram according to the
standard Szeg\"o Theorem.

\section{Acknowledgments}

We greatly benefited from multiple discussions with F. Essler, V.E. Korepin,
and B. McCoy.
AGA would like to acknowledge the financial support from Alfred P. Sloan foundation.




\begin{thebibliography}{00}




\bibitem{shiroishi}
	M. Shiroishi, M. Takahashi, and Y. Nishiyama, J. Phys. Soc. Jap.
	{\bf 70}, 3535 (2001).
\\ {\it Emptiness Formation Probability for the One-Dimensional
Isotropic XY Model.}

\bibitem{korepin93}
	V.E. Korepin, N.M. Bogoliubov, and A.G. Izergin, {\it Quantum
Inverse Scattering Mehod and Correlation Functions}, Cambridge
University Press, Cambridge, UK, 1993.

\bibitem{korepin94}
	V.E. Korepin, A.G. Izergin, F.H.L. Essler, and D.B. Uglov, Phys.
	Lett. {\bf A 190}, 182 (1994).
\\ {\it Correlation functions of the spin-1/2 XXX antiferromagnet.}

\bibitem{stroganov1}
	Yu. Stroganov, J. Phys. A-- Math. Gen. {\bf 34}, L179 (2001).
\\ {\it The Importance of being Odd.}
\\
	A. V. Razumov, and Yu. G. Stroganov, J. Phys. A-- Math. Gen. {\bf
        34}, 3185 (2001).
\\ {\it Spin chains and combinatorics.}

\bibitem{KMST_0-2002}
          N. Kitanine, J.M. Maillet, N.A. Slavnov and V. Terras, J.
          Nucl. Phys. {\bf B642}, 433-455 (2002).
\\ {\it Correlation functions of the  XXZ spin-$\frac{1}{2}$
Heisenberg chain at free fermion point from their multiple integral 
represenations}
\\
          N. Kitanine, J.M. Maillet, N.A. Slavnov and V. Terras, J.
          Phys. A: Math. Gen. {\bf 35}, L385-L388 (2002).
\\ {\it Emptiness formation probability of the XXZ spin-$\frac{1}{2}$
Heisenberg chain at $\Delta=\frac{1}{2}$.}

\bibitem{KMST_gen-2002}
	N. Kitanine, J.M. Maillet, N.A. Slavnov, and V. Terras, J. Phys. A:
	Math. Gen. \textbf{35}, L753-L758 (2002).
\\ {\it Large distance asymptotic behaviour of the emptiness 
formation probability of the XXZ spin-$\frac{1}{2}$ Heisenberg chain}

\bibitem{KLNS-2002}
          V.E. Korepin, S. Lukyanov, Y. Nishiyama and M. Shiroishi, 
          to be published in Phys. Lett. A,  
          arXiv:cond-mat/0210140.
\\ {\it Asymptotic Behavior of the Emptiness Formation Probability in
the Critical Phase of XXZ Spin Chain.}

\bibitem{abanovkor}
         A.G. Abanov and V.E. Korepin, Nucl. Phys. {\bf B 647}, 565,
	(2002).
\\ {\it On the probability of ferromagnetic strings in
antiferromagnetic spin chains.}

\bibitem{LSM-1961}
	E. Lieb, T. Schultz, and D. Mattis, Ann. of Phys. \textbf{16}, 407-466 
	(1961)
 \\ {\it Two Soluble Models of an Antiferromagnetic Chain}

\bibitem{mccoy}

	E. Barouch, and  B.M. McCoy, Phys. Rev. {\bf A 3}, 786 (1971).
\\ {\it Statistical Mechanics of the XY Model. II. Spin-Correlation
Functions}.



\bibitem{shrock}
	J. Kurmann, H. Thomas, and G. M\"uller, Physica A {\bf 112}, 235 (1982).
\\ {\it Antiferromagnetic Long-Range Order in the Anisotropic Quantum Spin Chain.}
\\
	G. M\"uller, and R.E. Shrock, Phys. Rev. {\bf B 32}, 5845 (1985).
\\ {\it Implications of direct-product ground states in the
one-dimensional quantum XYZ and XY spin chains.}

\bibitem{wu}
	T.T. Wu, Phys. Rev. {\bf 149}, 380 (1966).
\\ {\it Theory of Toeplitz Determinants and the Spin Correlations of the Two-Dimensional Ising Model. I}

\bibitem{FisherHartwig-1968}
	M.E. Fisher and R.E. Hartwig, Adv. Chem. Phys. {\bf 15}, 333 (1968).
 \\ {\it Toeplitz determinants, some applications, theorems and conjectures.}

\bibitem{widomsing}
	H. Widom, Amer. J. Math. {\bf 95}, 333 (1973).
\\ {\it Toeplitz Determinants with Singular Generating Functions.}

\bibitem{basor}
	E.L. Basor, and C.A. Tracy, Phys. {\bf A 177}, 167 (1991).
\\ {\it The Fisher-Hartwig conjecture and generalizations.}
\\
	E.L. Basor, and K.E. Morrison, Lin. Alg. App. {\bf 202}, 129 (1994).
\\ {\it The Fisher-Hartwig Conjecture and Toeplitz Eigenvalues.}

\bibitem{widomsupp}
	H. Widom, Ind. Univ. Math. J. {\bf 21}, 277 (1971).
\\ {\it The Strong Szeg\"o Limit Theorem for Circular Arcs.}

\bibitem{Ehrhardt-1997}
	T. Ehrhardt and B. Silbermann, J. Funct. Anal. {\bf 148}, 229-256 (1997).
\\ {\it Toeplitz Determinants with One Fisher-Hartwig Singularity.}

\bibitem{Ehrhardt-2001}
	T. Ehrhardt, Operator Th: Advances and App. {\bf 124}, 217-241 (2001).
\\ {\it A status report on the asymptotic behavior of Toeplitz determinants with Fisher-Hartwig singularities.}



\bibitem{AF}
	A.G. Abanov and F. Franchini, to be published.


\bibitem{dysmehta}
	J.des Cloizeaux and M.L. Mehta, J. Math. Phys. {\bf 14}, 1648 (1973).
\\ {\it Asymptotic Behavior of Spacing Distributions for Eigenvalues of Random Matrices.}
\\
	F. Dyson, Commun. Math. Phys. {\bf 47}, 171 (1976).
\\ {\it Fredholm Determinants and Inverse Scattering Problems.}




\end{thebibliography}
\end{document}